# The Privacy Coach: Supporting customer privacy in the Internet of Things

Gerben Broenink[1], Jaap-Henk Hoepman[1,2], Christian van 't Hof[3], Rob van Kranenburg[4], David Smits[5], Tijmen Wisman[6]

[1]TNO, the Netherlands
[2]Radboud University Nijmegen, Nijmegen, the Netherlands
[3]Rathenau Institute, The Hague, the Netherlands
[4]Fontys Applied Sciences, Tilburg, the Netherlands
[5]Mobi.Ubiq, Amsterdam, the Netherlands
[6]Independent consultant, the Netherlands

gerben.broenink@tno.nl, jhh@cs.ru.nl, c.vanthof@rathenau.nl, kranenbu@xs4all.nl, david.smits@mobiubiq.org, tijmenwisman@hotmail.com.

**Abstract.** The Privacy Coach is an application running on a mobile phone that supports customers in making privacy decisions when confronted with RFID tags. The approach we take to increase customer privacy is a radical departure from the mainstream research efforts that focus on implementing privacy enhancing technologies on the RFID tags themselves. Instead the Privacy Coach functions as a mediator between customer privacy preferences and corporate privacy policies, trying to find a match between the two, and informing the user of the outcome. In this paper we report on the architecture of the Privacy Coach, and show how it enables users to make informed privacy decisions in a user-friendly manner. We also spend considerable time to discuss lessons learnt and to describe future plans to further improve on the Privacy Coach concept.

## Introduction

Increasingly, products for sale in shops are being tagged by RFID tags. Also, more and more smart cards become contactless and are being used as travel passes, payment- or loyalty cards. This technology is similar to RFID. The tags attached to products contain a unique product- or item number, which can be read out wirelessly over a short distance by an RFID reader. Their function in shops and supermarkets is similar to the ubiquitous printed barcode. This also holds true for payment cards and travel passes (e.g. the OV chipcard in the Netherlands), where the unique number is not attached to a product but to a person. Using such a smart card means that one builds up an identity in a customer database, which records when, where one made which purchase or which trip. This raises serious concerns about the impact of RFID technology on privacy in our society. Who is allowed to gather what kind of information on you, and to what extent should you as a customer be informed about that? [Van 't Hof 2007]

That is why there are currently several initiatives [ISO] that try to inform consumers about the presence and use of RFID tags in the items that they buy or the cards they carry with them. The aim is to allow producers and shop owners to inform the consumer about what information they will collect, what they will do with it, and with whom they will share it. It is up to the consumer to decide whether to go ahead and buy the item, or not to buy the item after all. Consumers in these initiatives are informed through signs or logos in the shop or on the items themselves [Schermer 2009].

These approaches fail in two respects. First of all, the amount of information that can be communicated through a few logos is quite small, making it hard for the consumer to make an informed decision. Moreover, the number of logos producers are willing to put on an item is very small (hardly ever more than 1), meaning that current efforts do not go much further than informing the consumer about the mere presence of an RFID tag, without telling him much more. Secondly, the information is static. Once the logo is on the item, it cannot change. This either forces the producer to stick to the privacy policy (which is very inflexible) or to change the policy without being able to inform the consumer (which is bad for customer relations).

That is why we developed the Privacy Coach. It is an application running on the user's mobile phone, which is able to read the RFID tag of a product or an ID badge, and download the companion privacy policy. This privacy policy can then be compared with the user preferences, to decide whether the polciy of the product or badge matches with the user preferences. The phone should have a so-called Near Field Communication device so that it can communicate with RFID tags. Also, it should have an Internet connection. The application on the phone is supported by a database running on a server in the backoffice. A consumer sets his privacy preferences in a profile stored on his mobile phone. If he holds the phone close to a product in a shop or a smart card containing an RFID tag, the phone will read the tag number from the tag. It will then query (over the Internet, either through GPRS, UMTS or WiFi) the backoffice to retrieve the privacy policy corresponding to the tag number. It will then match the tag policy with the consumer privacy profile, and present the result of the match to the consumer on the display of the mobile phone in an intuitive and appealing manner.

This way, the Privacy Coach gives users information about and control over the privacy risks associated with the RFID tags they are surrounded with. Moreover, the privacy policies associated with RFID tags can be dynamic, offering a much more flexible scheme than logos alone.

The Privacy Coach is not a commercially available application. It was developed by members of DIFR (the Dutch Interdisciplinary Forum on RFID, http://www.difr.nl) as a proof of concept: to demonstrate that technologies already in place can be used in a different manner, to suit the needs of the user in stead of the needs of the supplier. The research and development effort reported in this paper was supported by the NLNet Foundation (http://www.nlnet.nl/). The software is open source. Details can be found on http://www.difr.nl/?page_id=10.

The remainder of this paper is organised as follows. We first discuss the relevance of privacy in the Internet of Things (IoT), and the particular paradigm of privacy protection we have adopted when developing the Privacy Coach. Then we describe the state of the art, followed by the system architecture (including a description of how a user interacts with the application). We finish with a discussion of lessons learnt when developing this application, and provide some conclusions based on our work.

## Privacy, the Coach, and the Internet of Things

In our daily lives we are increasingly tagged with smartcards: to enter the office, to identify ourselves at the fitness club or to pay in public transport. RFID tags are attached to objects surrounding us, or even the cloths we wear. These are elements of a control system, enforced by the owners of smart environments. The unique number on the card or the tag triggers events in databases, like: “you have paid, and are therefore granted access”. Meanwhile, such events are aggregated to build profiles in databases running in the backoffice: “this is a loyal customer”, or “this customer may be tempted to buy product X”. As a user, you can only see the responses of the devices, not the information processes running in the backoffice.

Legally, this practice is correct, as you have agreed with the privacy policy, as you have signed the User License Agreement when you purchased the transport pass or the loyalty card, probably without even reading it. In practice, this leads to information asymmetry: the provider of the products and services gains more information on you, while you hardly get any information on them in return. Moreover, there is no incentive for the provider to have a good privacy policy, as nobody reads such policies anyway [Van ‘t Hof 2007].

Spiekermann et al. [Spiekermann and Evdokimov, 2009] observe that although there are many protocols and proposals for limiting access to RFID tags (either by killing them completely or by requiring the reader to authenticate), few systems have been proposed that allow effective and fine grained control and management over access permissions. They argue that much more research is needed in so called user schemes, where users exert immediate control over their RFID tags.

### Agency to increase privacy

Web 2.0 is becoming mundane for the younger generations: people routinely build informal layers (networks) between formal institutions and their everyday practices. As of yet there are no people driven examples (although there are attempts like Touchatag[1]) in the IoT, to show the benefits of smart objects, smart homes, streets and cities. The key to facilitating the Internet of Things for citizens is to realize that they (the citizens) are not stand-alone subjects anymore. Instead, based on their Web 2.0 experience, they are always connected to other individuals and communities.

This entails that the IoT should allow for end-user programming, providing the agency of users to tweak, and that its infrastructure is generic and designed to be as open as possible to allow for emerging services and applications. The IoT should be designed for a generation that is not only hungry for content and services, but also wants to modify and personalise the tools that they use to sculpt their identity. Privacy in this generation should be modelled as ‘privacies’: different datasets that are tied to different actualizations of one's identity. Some data are essential in one actualisation of one's identity, not in others.

[1] www.touchatag.com

This paradigm shift from privacy to privacies acknowledges that in a hybrid environment we leave different traces and might want to build temporary actualizations of personalities around these traces, not exposing our entire personality all the time.

This is why the Privacy Coach turns the tables around: you as a customer use your mobile phone to scan the smartcard or the RFID tag, and the coach tells you whether the corresponding privacy policy fits your privacy preference. This privacy preference is the result of a questionnaire you filled in when you started using the coach and is stored on the mobile phone. For example: "the provider can register whether your card has entered their facility, but they cannot connect it to your name". Or: "you agree they build up a customer profile on your purchases, but only if that results in certain discounts". Next time you are offered a new RFID smart card, just hold it to your phone, which states "match" or "no match". If the card does not match your profile, you can ask the provider what he is up to with your personal information. If it does match, you can be at ease. The Privacy Coach provides customers agency over their privacy.

The privacy policies associated with a tag or a smart card, are retrieved from a central database. The policies stored in this database can come from several sources. In fact, the system could easily use several different databases maintained by different organisations. The primary source for the privacy profiles are the businesses themselves. One could wonder whether providers of smart cards and tags are willing to go through this effort. On the one hand, they are obliged by law to inform customers what they do with their personal information, resulting in the agreements described above. Providers who care about the privacy of their customers, or perceive their privacy policy as a unique selling point, may be tempted to use the Privacy Coach procedure. On the other hand, independent organisations like consumer rights organisations, could collect and offer such privacy policies, increasing the likelihood that the information in these policies is correct and not misleading. This is another application of the observation that citizens are not stand-alone subjects anymore, but have many relationships with other individuals and other communities.

The Privacy Coach is a concrete application, by having privacy levels on your mobile phone. In industry terms you have an identity manager, in privacy activist terms you would have the equivalent of Melanie Rieback's RFID Guardian – a firewall.

## State of the art

Proper privacy protection within RFID based systems is of paramount importance [Spiekermann and Evdokimov, 2009, Garfinkel et. al. 2005]. In the academic literature, the emphasis is on implementing privacy enhancing technology on the RFID tags themselves [Juels 2006., Hoepman and Joosten 2009]. The Privacy Coach is based on a different approach.

One of the sources from which the concept of the Privacy Coach was derived is our experience in Japan, where one can see a quite different approach to using RFID in public space. While it is customary to setup a RFID system in such a way that people and products are labelled with tags which are read by the environment, Ken Sakamura of the Japanese Ubiquitous Laboratory turns it around: people read tags in their environment. His lab developed several trials to make the Tokyo metropolitan a Ubiquitous city which can be read with a hand held RFID reader with internet connection, the so called Ubiquitous Communicator (UC). In the Tokyo Midtown area for example, one can do a Ubiquitous Art Tour. The reader senses tags in sculptures that trigger information on demand: an explanation of the art piece, a movie on how it was made etc. In the Ginza shopping area, the UC provides its user with shopping information: discounts, directions, etc. In the zoo, the UC provides visitors with information on the animals. The principle is quite simple: the tag in the environment is nothing more than a unique code (128 bits) which refers to information in the U-Code database, which is openly accessible through the internet. [Schilpzand and Van ‘t Hof 2008]

The RFID Guardian [Rieback et al, 2006] was another source of inspiration. The Guardian is also a portable device, specifically designed for its purpose. The main idea is to jam all reader to tag communication, except for reader requests that satisfy a pre-defined privacy rule. In that sense the RFID Guardian operates like a firewall between tags and readers. This approach has its own shortcomings. For one, it is extremely hard to ensure that all reader to tag communication is effectively blocked in all cases. Moreover, tags themselves are not protected at all, leaving them vulnerable when the Guardian is out of range or malfunctioning.

## System Architecture

The system architecture consists of client software running on a mobile device and a central Policy Provider server where RFID tags numbers and privacy policies are stored (see Figure 1). The user can access privacy policies, based on the ID or number of the tag, which is registered in the Policy Provider. The mobile device transmits the tag-number via wireless Internet (GPRS, UMTS, WIFI) and the server returns the requested policy. The received privacy policy is then matched on the mobile device with the stored personal privacy settings and the mobile device displays the result of the policy comparison. This results in either a match or a non-match. In case of a non-match the privacy policy of the tag does not comply with the stored privacy preferences of the user. Feedback on the occurrence of a miss-match in the policy is reported back to the user in order to make the final decision for using the tag.

To read RFID tags mobile devices with NFC capability are used. The system works with the three NFC mobile phones that are currently available on the market:

- Nokia 6131 NFC
- Nokia 6212 NFC
- Nokia 6216

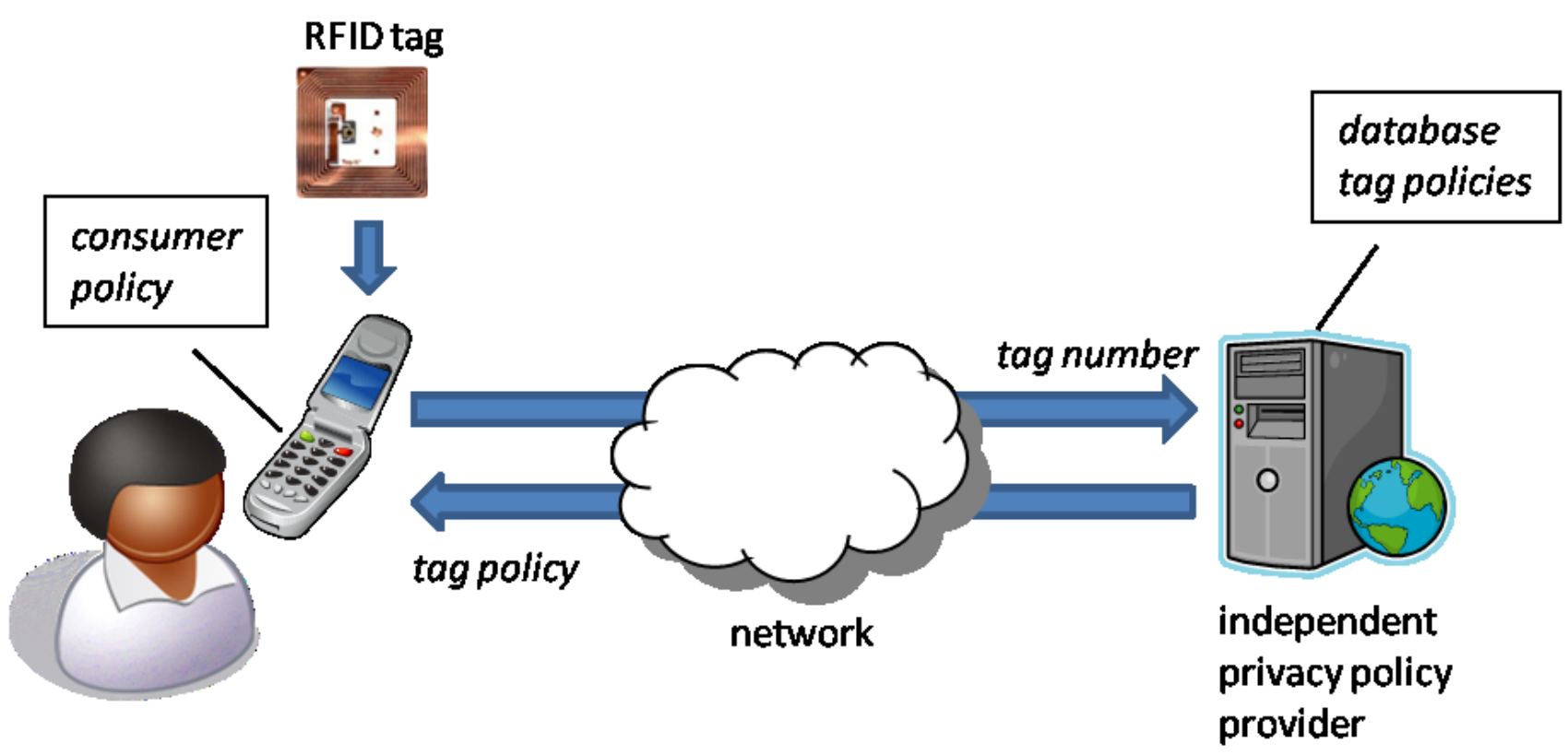

**Figure 1 Privacy Coach system architecture**

We note that due to incompatibilities with the NFC and Electronic Product Code (EPC) standards (the former operates on the 13.56 MHz (HF) frequency band and the latter on the 860 MHz – 960 MHz (UHF) frequency band), such phones cannot read EPC based tags. In our demonstrator we used Mifare based tags instead.

### Privacy profiles

We distinguish two different kinds of RFID tags:

1. ID Badges
2. RFID tags on products.

For both types of RFID chips we have created separate privacy policy profiles. These privacy policies are stored on the mobile device and the attached privacy policy of the RFID tag is stored on the server.

*ID Badges*

ID Badges are identity cards that can be used to access buildings or services, for payment purposes or as a customer loyalty card. In the following chart (Figure 2) a diagram of the implemented privacy profile is shown. Based on the use of the card additional information or promotions can be offered to the user in the form of e-mail, SMS text messaging or on displays on the premises or at the cashier desk.

*RFID tags on products*

RFID tags on products are used by retailers for supply chain management. In stores these tags can be used to provide pricing of products and for additional information about the product. In the following diagram (Figure 3) the privacy profile for products with RFID tags is shown. Also additional promotions can be offered based on the profile of the user.

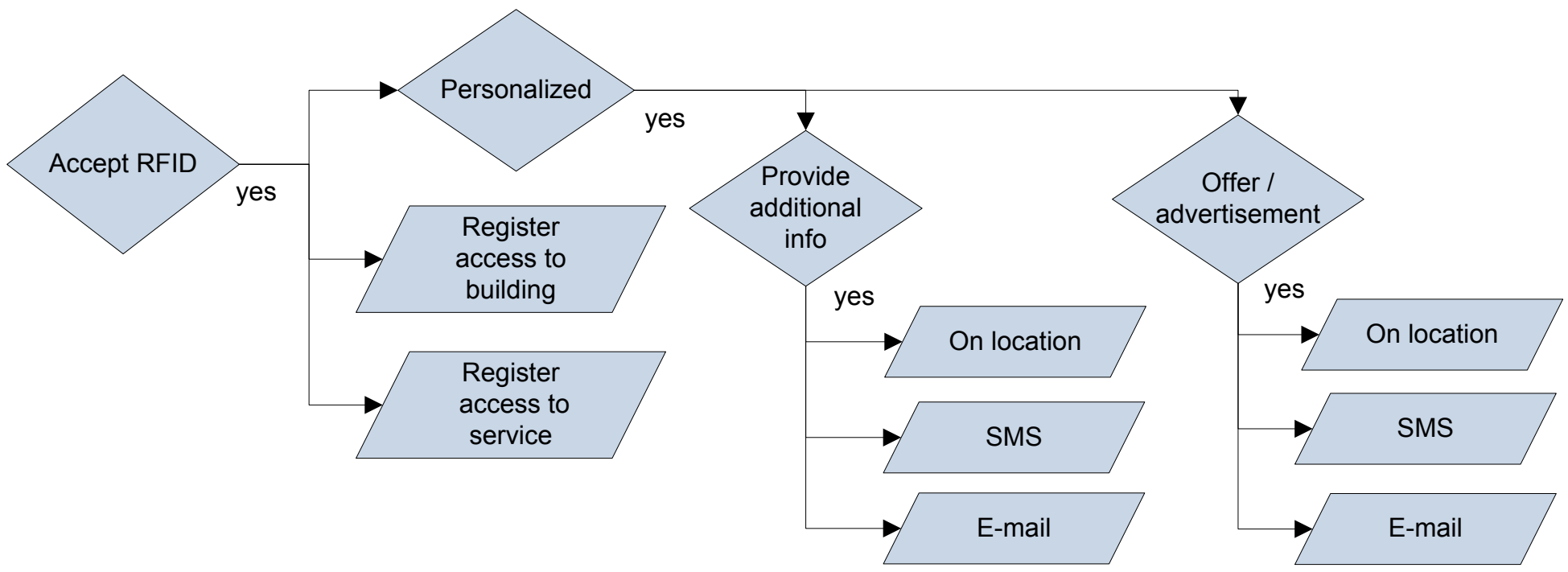

**Figure 2 Privacy profile for ID Badges**

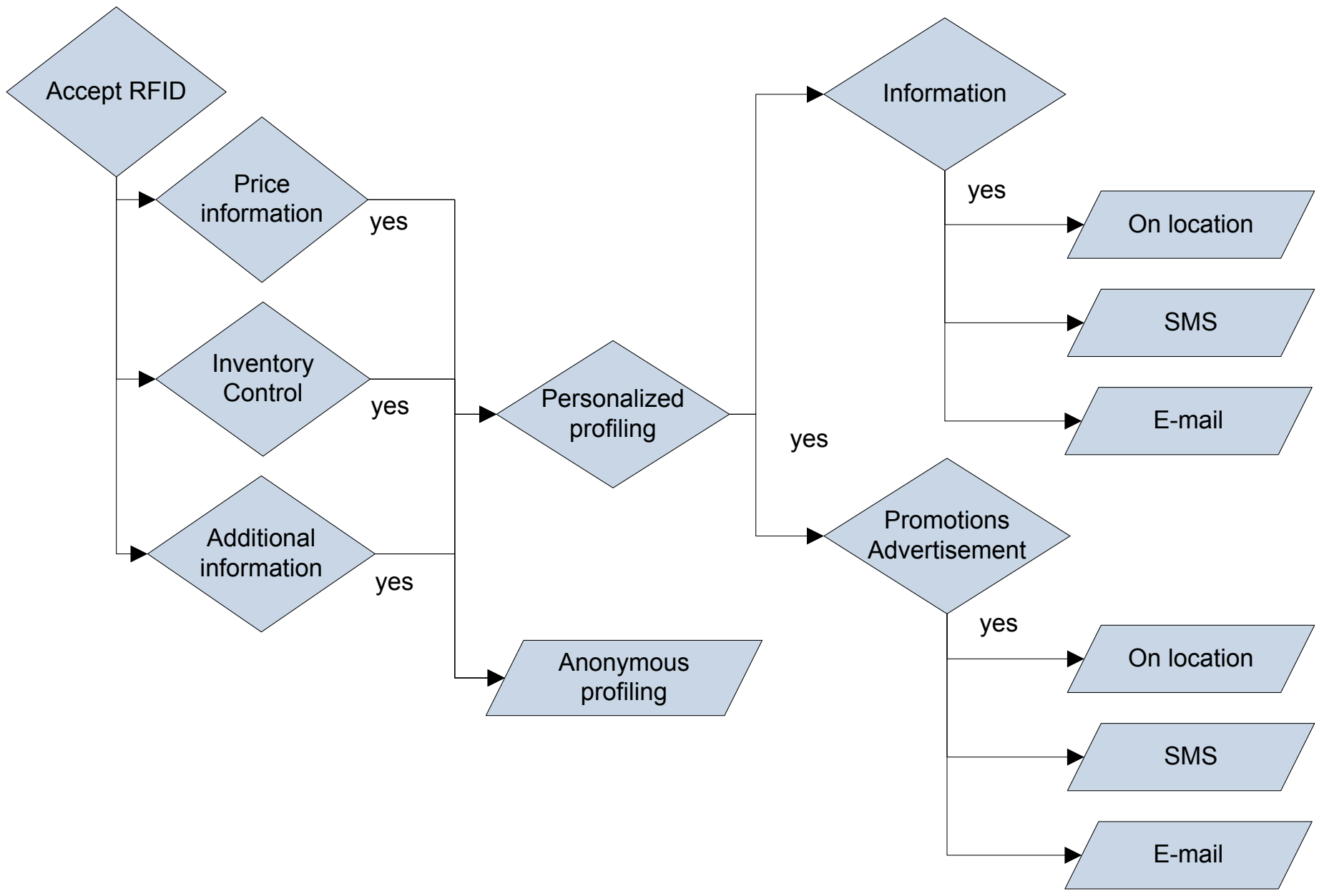

**Figure 3 Privacy profile for RFID Tags**

## The mobile client

The mobile client offers three different functions.

1. Configure privacy policy user;
2. Read RFID tag and policy;
3. Match user and tag policy.

*Configuring privacy policy*

The privacy preferences of the user are requested the first time that the client application is run. These preferences are requested through a question and answer wizard, following closely the profiles given above. The privacy profile of the user is constructed based on the given answers and is stored on the phone.

*Reading RFID chips and requesting tag policy*
When the user would like to see whether he or she can accept a product or ID badge this can be achieved by touching the RFID tag with the mobile phone. The phone will read the tag number, which identifies the product or badge. This tag number is used to request the associated policy.

*Comparing policies*
When the policy of the RFID tag is received, the client matches this policy with the stored policy on the mobile device. The application will check each variable of the policies. If one of the variables in the policy is not acceptable to the user, the application will display a 'no match' screen. When all variables are positively matched, the application will display a 'match' screen.

### The server

The Privacy Policy provider is a server connected to the Internet. Privacy policies are requested through a publicly available web service. This web service returns the policy in a fixed XML format that can be interpreted by the client on the mobile phone. The server does not store any user data and with the request from the mobile phone only the tag number is transmitted.

Example request:
http://mrenne.s42.eatj.com/PrivacyCoachService/tagpolicy/e4f6060c

The server can register tag numbers through a web interface that is available on the server. Alternatively tag numbers can be registered via a web service interface and in this way creating possibilities to automatically register tag numbers from within other systems.

## Lessons learned and future plans

As explained in the system architecture section, NFC is not RFID: NFC phones can read Mifare tags, but not EPC global tags. To overcome this limitation a Bluetooth RFID reader could be used together with a mobile phone.

The request form the mobile phone to the Service Provider is done anonymously and only the tag-number is send. The matching is done on the mobile device and the decision logic is on the mobile phone. This makes it more difficult to change or extend the decision logic, the privacy policy language and the matching algorithms. Al-ternatively the matching of the privacy profiles could be done server side, but this raises certain privacy concerns.

The concept of the Privacy Coach could be extended to include other (non RFID) application areas, like software applications, or websites, that have similar unreadable User License Agreements that none ever reads. A Privacy Coach 2.0 could be used, saving the customer the effort of reading these texts and giving more honest providers a chance to distinguish themselves from others.

Similarly, the model of the Privacy Coach can be taken a step further towards a Matching Coach. Similar to matching profiles on dating sites, customers can set their personal profile, using the Matching Coach to see what kind of services or products fit their needs. Examples of applications where such a Matching Coach could be useful are matching food products to diets (allergies, weight loss), negotiating service levels and corresponding discounts, or buying products depending on the ethical standards of production (e.g. ecological food, carbon-neutral production, etc.).

The current model of the Privacy Coach is static and the privacy policies and profiles are rather crude. No consideration is given to the context in which certain personal information is used or collected. Moreover, if your privacy profile does not match the privacy policy, your only option is not to accept the tag (or to accept the policy, of course). It would be much more flexible if the policy would be negotiable, and could be adapted to certain contexts. For example, it makes a big difference whether information is being collected and stored to offer products, or to offer current delays of the train one wants to take. Still both types of information could be send by a public transport company. So in an ideal world the consumer can determine why information is being stored, thereby ensuring the type of information that one receives. Another factor that is important in determining the consumers preferences is the length of the storage of information. In an ideal world one can also determine this specific aspect. Also a consumer could be allowed to determine whether he wants to use a service anonymously. Instead of being linked to the consumers real identity, the information could be linked to an pseudonym. The list above is not exhaustive, there could very well be more conditions that are worth negotiating.

If consumer preferences and business policies do not match initially, the Privacy Coach could become a mediator that facilitates the negotiation to align business policies and user preferences. This is quite ambitious though, because it means each individual tag may have a special privacy policy associated with it. This has to be managed to ensure that all business processes that treat a particular RFID tag abide the particular privacy policy at hand.

## Conclusions

We can envisage a definition of IoT identity as an ever changing mix of relations between a physical body of a person, his or her objects and a ‘smart’ environment. All current computing paradigms put connectivity and content centric networking central: Internet of Things, Pervasive Computing, Ubicomp, Ambient Intelligence; the environment becomes the interface. The connectivity will be an ecology of RFID, active sensors, biometrically related smart camera data, 2D barcodes and 6LoWPAN: IPv6 over Low-Power wireless Area Networks. Monitoring mechanisms will be build into devices themselves: “if a guest is charging their electric car at a friends house, we should consider applications that will understand that the charge should appear on the guests electric bill and not that of the friend." It is unproductive to attempt to isolate old constants in such an environment. The privacy of objects is just as relevant or irrelevant as the privacy of persons in this fluid ecology that is called ‘identity’.

With the Privacy Coach we have shown that existing technology can be used in a different manner, to fit this vision, and to suit the needs of the user instead of the supplier. When users are enabled to read their environment, instead of the environment reading them, we can resolve the information asymmetries that currently hamper the Internet of Things.